\documentclass[twocolumn,amsmath, amssymb, aps, pra,reprint, longbibliography,nofloatfix,nofootinbib,superscriptaddress]{revtex4}
\pdfoutput=1

\usepackage[utf8]{inputenc}
\usepackage[urlcolor=blue, colorlinks=true,citecolor=blue]{hyperref}

\usepackage[english]{babel}
\usepackage{natbib}

\usepackage{amsfonts}
\usepackage{physics}
\usepackage{graphicx}

\usepackage{graphicx}
\usepackage{amsmath,bm}

\usepackage[normalem]{ulem}
\usepackage{bbold}
\usepackage{dsfont}
\usepackage{xcolor}
\usepackage{float}

\newcommand{\ii}{\mathrm{i}}

\begin{document}
\title{Information plane and compression-gnostic feedback in quantum machine learning.}

\author{Nathan Haboury}
\affiliation{Terra Quantum AG, Kornhausstrasse 25, 9000 St.~Gallen, Switzerland}

\author{Mo Kordzanganeh}
\affiliation{Terra Quantum AG, Kornhausstrasse 25, 9000 St.~Gallen, Switzerland}

\author{Alexey Melnikov}
\affiliation{Terra Quantum AG, Kornhausstrasse 25, 9000 St.~Gallen, Switzerland}

\author{Pavel Sekatski}
\affiliation{Terra Quantum AG, Kornhausstrasse 25, 9000 St.~Gallen, Switzerland}

\begin{abstract}
The information plane~\cite{tishby2000information,shwartzziv2017opening} has been proposed as an analytical tool for studying the learning dynamics of neural networks. It provides quantitative insight on how the model approaches the learned state by approximating a minimal sufficient statistics. In this paper we extend this tool to the domain of quantum learning models. In a second step, we study how the insight on how much the model compresses the input data (provided by the information plane) can be used to improve a learning algorithm.
Specifically, we consider two ways to do so: via a multiplicative regularization of the loss function, or with a compression-gnostic scheduler of the learning rate (for algorithms based on gradient descent). Both ways turn out to be equivalent in our implementation.
Finally, we benchmark the proposed learning algorithms on several classification and regression tasks using variational quantum circuits. The results demonstrate an improvement in test accuracy and convergence speed for both synthetic and real-world datasets. Additionally, with one example we analyzed the impact of the proposed modifications on the performances of neural networks in a classification task. 
\end{abstract}

\maketitle

\section{Introduction}

Machine learning has achieved remarkable success over the past decades, to the extent that it has sparked both enthusiasm and concern among influential thinkers \cite{de_miranda_efficient_2019}. The ability of machine learning models to learn from vast amounts of data and make accurate predictions has revolutionized industries ranging from healthcare to finance. However, as these models grow increasingly complex, they also demand more computational power, pushing the limits of classical computing \cite{vegh2018statistical}. This has led to the exploration of new paradigms, including quantum computing, which promises to redefine the boundaries of what is computationally possible. \\

Quantum machine learning (QML) offers the potential to leverage the unique capabilities of quantum hardware to enhance machine learning processes~\cite{feynman_quantum_1986,Davide_Castelvecchi_2023,acharya_suppressing_2023,qml_review_2023}. While there is broad consensus that QML and variational quantum algorithms could be transformative for inherently quantum problems such as those in chemistry \cite{51309}, material science \cite{babbush2023exponential,arute2019quantum}, and particle physics \cite{Blance_2021,Wu_2021,di_meglio_alberto_2021_5553775}, the impact of QML on classical problems remains an open question \cite{JADHAV20232612,vasques_application_2023}. However, it has been demonstrated that QML can provide up to exponential learning speedups on specially designed datasets \cite{servedio2004equivalences, jerbi2021parametrized, sweke2021quantum, liu2021rigorous, yamasaki2023advantage, molteni2024exponential}, and it is also understood that quantum circuits allow sampling from a broader class of probability distributions compared to classical algorithms~\cite{Hangleiter2023computational}. Nevertheless, as is often the case in machine learning, the real test QML’s usefulness will come from future practical applications~\cite{rainjonneau2023quantum,sagingalieva2023hyperparameter,sagingalieva2023hybrid,sedykh2024hybrid}. \\

In this work, we focus on taking a concept originally developed in classical machine learning and exploring its application within the realm of quantum machine learning. Our goal is to adapt and extend this concept to determine its effectiveness and potential advantages in a quantum context. By building on recent advances in the field, we aim to contribute to the development of quantum machine learning algorithms that are theoretically robust and capable of offering practical benefits.

\section{Background} \label{sec:background}

\subsection{Minimal sufficient statistics and the information plane}

In a supervised machine learning problem the goal is to find the function $f$
that infers the value of the variable $Y$ (the label) encoded in the input $X$ (the data) with high accuracy $f(X)\approx Y$~\cite{Goodfellow-et-al-2016}. In other words, $f$ should distill a compact representation of data that retains vital information on the label while discarding all irrelevant details. For a labeled data set $\{(x_i,y_i)\}$ this task can be viewed as approximating the \textit{minimal sufficient statistics} $f_*$~ \cite{tishby2000information, SHAMIR20102696}, that is a function 
\begin{equation}
f_*: X\mapsto T=f_*(X)
\end{equation}
that is sufficient, i.e. $p(Y|X)=p(Y|T)$, and minimal, i.e. $T$ can be computed from any other sufficient statistics~\cite{fisher1922mathematical}.\\

Of course, in order to avoid over-fitting one only aims to approximate $f_*$ to a certain degree, which is in theory dictated by the Vapnik–Chervonenkis dimension of the underlying model class~\cite{shalev2014understanding,mohri2018foundations}. Nevertheless, the learning process can be analyzed through the convergence of the model to the minimal sufficient statistics. In turn, this convergence can be captured by the dynamics of the mutual information between the input data $X$, the label $Y$, and the compressed representation of data $T=f(X)$ propagating through the model, e.g. the activation of the neurons in a given layer in the case of neural networks~\cite{shwartzziv2017opening}. Concretely, for random variables $X$, $Y$, and $T$\footnote{In most cases it is natural to take the random variable $X$ to be uniformly distributed on the training data set.} the mutual information 
\begin{equation}\label{eq: MI}
I(X:T)= I(T:X) = H(X) + H(T) - H(X,T),
\end{equation}
with the Shanon entropy $H(X)= -\mathds{E}\big(\log p(X)\big)$ \cite{6773024}, is a measure of correlations between two random variables \cite{1057418, cover_elements_nodate}. It describes how much information does one variable carry about the other~\cite{cover_elements_nodate}. Now, for a \textit{sufficient} statistics the data processing inequality $I(X:Y)\geq I(T:Y)$ is saturated.
While a \textit{minimal} sufficient statistics shall minimize $I(T:X)$ in addition. \\

In \cite{shwartzziv2017opening} Shwartz-Ziv and Tishby analyzed the learning process in (deep) neural networks via the \textit{information plane} defined by the quantities $I(T:X)$ and $I(T:Y)$, by looking at how these two quantities evolve during learning. In particular,  they identified two learning phases. A short fitting phase where both $I(T:X)$ and $I(T:Y)$ increase and the model learns to represent the input data faithfully, followed by a long compression phase where $I(T:X)$ decreases dramatically while $I(T:Y)$ remains close to its past maximal value. Another interesting aspect is the evolution of $I(T:X)$ through the subsequent layers of a trained neural network. By data processing inequality it decreases from layer to layer, but moreover, it can decrease by a factor between each layer, which shines some light on the success of deep neural networks as efficient compressors. While the universality of these conclusion was later questioned~\cite{saxe2019information}, the information plane remains a useful tool to study the learning process.
\\

\subsection{Taming continuous data representations}

An important subtlety that we have so far completely ignored comes from the fact that the data is commonly processed into continuous (up to machine precision) real variables $T$, like neuron activation. Then, different data points $x_i$ will in general be deterministically processed into different real vectors $t_i$. In this case, since the mutual information does not change under reversible processing, neither $I(T:X)$ nor $I(T:Y)$ will change during learning, albeit the neural network may learn to separate the data subsets associated with different labels (in a classification task), see~\cite{amjad2019learning} for a detailed discussion. An intuitive solution is not to take the quantities  $I(T:X)$ and $I(T:Y)$ literally. Rather one thinks of them as some approximations of the mutual information that can be computed efficiently and have a nontrivial learning dynamics. Various ways approximate the mutual information or compute different quantities with a similar interpretation have have been proposed in the literature, see e.g. \cite{alemi2016deep,chalk2016relevant,shwartzziv2017opening,kolchinsky2019nonlinear,amjad2019learning,cvitkovic2019minimal}.

A simple and intuitive way to implement such approximations~\cite{shwartzziv2017opening} is to discretize the values of $t_i$ to some empirically chosen precision . Formally this corresponds to applying a map of the form 
\begin{align} \label{eq: disc map}
\mathcal{B} : \mathds{R}^M &\to [1,b]^M 
\end{align}
to all the real vectors $t_i$. Albeit it has some drawbacks (see below), we will adapt a similar approach when discussing quantum learning models in Sec.~\ref{sec:inf QML}.

\subsection{The information bottleneck method}

So far we have discussed the information plane and its variants as a lens to look at the learning dynamics of neural networks (which are trained independently). Nevertheless, it was originally introduced~\cite{tishby2000information} and mostly discussed in various variants~\cite{tishby2015deeplearninginformationbottleneck,chalk2016relevant, strouse2017deterministic,kolchinsky2019nonlinear,amjad2019learning,cvitkovic2019minimal} as a learning method on its own. The basics idea~\cite{tishby2000information} is to devise a learning algorithm that minimizes the information bottleneck loss functions
\begin{equation}\label{eq: LIB}
L_{\rm IB} = I(T:X) - \beta I(T:Y),
\end{equation}
where the hyper-parameters $\beta>0$ controls the importance of the compression (low $I(T:X)$) of the input data vs the retention of useful information (high $I(T:Y)$). Note that here the presence of the negative term $\beta I(T:Y)$ ensures that a trivial compressor discarding all input data is suboptimal.

In this paper, we are not going to follow this path. Instead of considering the information bottleneck as the central quantity that should be minimized by the algorithm, in Sec.~\ref{sec:problem_setting} we ask how monitoring the evolution of $I(T:X)$, as a proxy for data compression, can be used to improve a given learning algorithm. In our case the term $I(T:Y)$ becomes obsolete, since the original learning algorithm  encloses a loss function to quantify the performance of model's current state.

\section{The information plane in quantum models}
 \label{sec:inf QML}

In quantum machine learning the data $X$ is processed on quantum hardware. It is now carried by a quantum system $Q$ with the associated Hilbert space $\mathcal{H}_Q$ of dimension $d$, usually several qubits, all the way until the final measurement. Hence, the pairs of variable $Q$ with $X$ and $Q$ and $Y$ are now be described by quantum-classical states\footnote{We here consider the learning on classical data, albeit a similar discussion can be made for quantum data where $X$ is also carried by a quantum system.} 
\begin{equation}\label{eq: quantum state} \begin{split}
\rho_{XQ} &= \frac{1}{|\{x_i\}|} \sum_{i}\ketbra{x_i}_X \otimes \rho^{(i)}_Q, \qquad\\
\rho_{YQ} &= \frac{1}{|\{x_i\}|} \sum_{i}\ketbra{y_i}_Y \otimes \rho^{(i)}_Q
\end{split}\end{equation} rather then joint probability distributions $p(XT)$ and $p(YT)$ from the last section. Here, $\rho^{(i)}_Q$ is the global quantum state of all the qubits \textit{at some point in the circuit}, it depends both on the input $x_i$ and the variational parameters (see section~\ref{sec:Implementation on synthetic dataset} for a concrete example with a specific circuit architecture.) The case of quantum hardware exhibits a few important differences from the classical models that are worth pointing out. \\

First, in contrast to NN where all of the data is uploaded in the first layer, for quantum circuits it is common to use architectures with \textit{data re-uploading}~\cite{P_rez_Salinas_2020}. In this case, there is no prior reason that the information that $Q$ carries on $X$ or $Y$ decreases until the last re-uploading layer of the circuit is passed.\\

Second, even after the last re-uploading layers, the dynamics of the \textit{global state} of all qubits is (ideally) unitary and thus invertible. Hence, when looking for compression, instead of the global state, it can be insightful to consider the marginal state of a subset of qubits or the distribution of an observable on those qubits, typically the one used for the final readout. To account for the first possibility we split the qubits in two groups $Q = Q' J$, the target and the junk defining the partition of the Hilbert space $\mathcal{H}_{Q}= \mathcal{H}_{Q'}\otimes \mathcal{H}_{J}$ and simply discard the junk system 
\begin{equation}
    \rho_{Q'}= \tr_{J} \rho_{Q'J}.
\end{equation}
For the second possibility, given an observable described by a Hermitian operator $\hat M=\hat M^\dag$, we introduce the twirling channel
\begin{equation}\label{eq: dephasing}
\mathcal{D}_{\hat M} : \rho \to \mathcal{D}_{\hat M}(\rho) = \sum_m \ketbra{m} \rho \ketbra{m},
\end{equation}
where $\ket{m}$ are the eigenstates of $\hat M$.
\\

 Finally, the third major difference with NNs is the fact that once it has been uploaded in the circuit the information processed by the model is \textit{quantum}.
In particular, it can not be copied or broadcasted like the activation values of neurons. For the same reason the quantum mutual information 
\begin{equation}\label{eq: QMI}
    I(X:Q)_{\rho_{XQ}} = H\left( \frac{1}{|\{x_i\}|} \sum_{i}\rho^{(i)}_Q\right) - \frac{1}{|\{x_i\}|} \sum_{i} H( \rho^{(i)}_Q),
\end{equation}
with the von Neumann entropy $H(\rho) = - \tr \rho \log \rho$, increases if we obtain more copies of the quantum state $\rho_Q^{(i)}\to (\rho_Q^{(i)})^{\otimes s}$, by e.g. sampling the model several times for the same input data. The quantity $I(X:Q)_{\rho_{XQ}}$ in Eq.~\eqref {eq: QMI} is thus not suitable to analyze how the model compresses information since we do not care about the sampling cost of retrieving this information (provided it is not forbidding). A drastic way to avoid any sampling considerations is to formally introduce a "tomography" map 
\begin{equation}
\mathcal{T} : \rho \mapsto t =(\rho_{11},\text{Re}(\rho_{1d}),\text{Im}(\rho_{1d}),\dots \rho_{dd})
\end{equation}
which takes a quantum system $Q$ and outputs a classical description of its density matrix (in the computational basis). The variable $t$ here is carried by a classical register $T$ containing the $d^2-1$ real parameters needed to specify $\rho_Q$. The map $\mathcal{T}$ is of course non-physical, nevertheless, it can be approximated to arbitrary precision by the standard quantum state tomography procedure given enough uses of the quantum hardware. Formally this can be summarized by the inequalities $I(T:X)_{\rho_{X \mathcal{T}(Q)}}\geq I(Q:X)_{\rho_{XQ^s}} \geq I(Q:X)_{\rho_{XQ}}$ (same for $Y$), where the first inequality is saturated in the asymptotic sampling limit $s\to \infty$ with the underlying state reading $\rho_{XQ^s}=\frac{1}{|\{x_i\}|} \sum_{i}\ketbra{x_i}_X \otimes (\rho^{(i)}_Q)^{\otimes s}$.\\

At this point, $T$ is a classical random variable and we recover the setting discussed in the previous section, with the information plane defined by the quantities $I(T:X)$ and $I(T:Y)$.
 This runs into the same "continuous data representation" problem as for neuron activation -- in general different data points $x_i$ result in different $t_i$ for all learning steps, rendering the mutual information $I(T:X)$ trivial. As already mentioned, a possible simple way to solve this issue is to discretize the state vectors $t_i$ with the map $\mathcal{B}$ in Eq.~\eqref{eq: disc map} to an empirically chosen precision, see below for examples. 

In the examples of Sec.~\ref{sec:results} we are going to use this discretization of the wave-vector trick to approximate $I(T:X)$.
Speaking very loosely, the information encoded in the discretization state vector $\mathcal{B}(t_i)$ can be thought of as having access to some number of copies of the quantum state $\rho^{(i)}_Q$. As an outlook, we believe that it is interesting to think about other methods to quantify the level of compression achieved by quantum models. An obvious possibility is to apply the other existing methods~\cite{alemi2016deep,chalk2016relevant,kolchinsky2019nonlinear,amjad2019learning,cvitkovic2019minimal} to the classical description of the wave-vector $t=\mathcal{T}(\rho)$. However one can also think of natively quantum possibilities, e.g. the $s$-copy quantum mutual information  $I(Q:X)_{\rho_{XQ^s}}$ for some choice of $s$.

\subsection{Training a PQC and the information plane: an example}

Before going any further, let us illustrate the introduced concepts 
on the example of the four-qubit parameterized quantum circuit (PQC) depicted in Fig.~\ref{fig:Full_circuit}. The setting is discussed in Sec.~\ref{sec:results} in full detail. For now, it is only important to know that it is trained to perform a binary classification task $(y_i=\pm 1)$ on a synthetic dataset of size $n$, and that this is done by measuring the first qubit in the computational basis and using the sign of the expected value of the corresponding observable to estimate the label $\hat y_i =  \text{sign} \left(\tr \sigma_Z^{(1)}\rho_Q^{(i)}\right)$. For training we use the standard mean squared error loss function
\begin{equation}\label{eq: MSE loss}
     L_{\text{MSE}}=\mathds{E}[(Y-\hat Y)^2] ={\frac {1}{n}}\sum _{i=1}^{n}\left(y_{i}-{\hat {y}_{i}}\right)^{2},
\end{equation}
and simulate the PQC on QMware~\cite{perelshtein2022practical} using Pennylane~\cite{bergholm2018pennylane,kordzanganeh2023benchmarking}.\\

To understand the learning dynamics of our model and where the compression happens, we consider the following variables
\begin{itemize}
\item[\textit{(i)}] For $\rho_Q$ denoting the full state of the four qubits $Q=Q_1Q_2Q_3Q_4$ after the last layer of the circuit, we define the variable 
\begin{equation}
     T_\text{all} = \mathcal{T}(\rho_Q).
\end{equation}
\item[\textit{(ii)}] For $\rho_{Q_1}$ denoting  the marginal state of the  first qubit $\rho_{Q_1}= \tr_{Q_2,Q_3,Q_4} \rho_Q$  after the last layer of the circuit, we define the variable 
\begin{equation}\label{eq: Tall}
     T_1 = \mathcal{T}(\rho_{Q_1}).
\end{equation}
\item[\textit{(iii)}] For the dephasing map $\mathcal{D}_{\sigma_Z}$  in Eq.~\eqref{eq: dephasing}, that can also be written as
$\mathcal{D}_{\sigma_Z}(\rho_{Q_1}) = \frac{1}{2}\left(\rho_{Q_1} + \sigma_Z \rho_{Q_1} \sigma_Z \right)$
 the tomography map defines the variable 
\begin{equation}\label{eq: T1z}
      \mathcal{T}\big( \mathcal{D}_{\sigma_Z}(\rho_{Q_1})\big) = (\rho_{11},\rho_{22},0,\dots,0)
\end{equation}
which is equivalent to the expected value of the $\sigma_Z$ observable on the first qubit. In this setting, we thus consider the variable
\begin{equation}
    T_1^Z = \tr \rho_{Q_1} \sigma_Z = \rho_{11}-\rho_{22}
\end{equation}
\end{itemize}
The variable $T_\text{all}$ and $T_{1}$ represent the information contained in all, respectively the first qubit, after the last layer of the circuit. The variable $T^Z_{1}$ represents the information encoded in the distribution of the $\sigma_Z$ observable (used for classification). 

\begin{figure}[ht]
    \centering
    \includegraphics[width=\columnwidth]{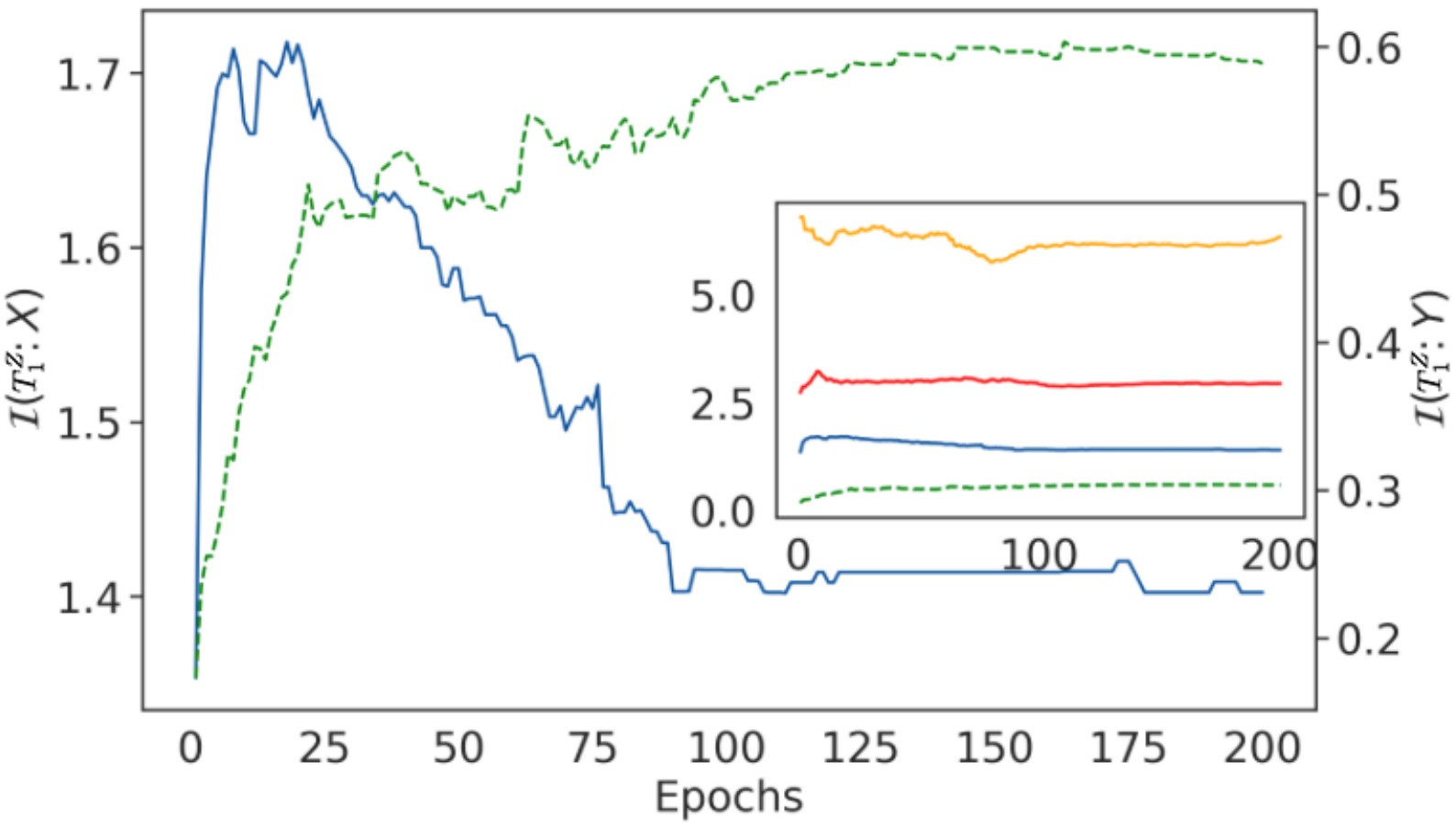}
    \caption{
    \textbf{(Main)} The information on the label $I(T_{1}^Z:Y)$ (dashed line) and on the data $I(T_{1}^Z:X)$ (full line) encoded in the distribution of the $\sigma_Z$ observable on the first qubit after the last layer, see~Eq.~\eqref{eq: T1z}. \textbf{(Inset)} From bottom to top: $I(T_\text{all}:X), I(T_{1}:X)$ and $I(T_1^Z:X)$ -- the information on the data encoded by all the qubits, the first qubits, and the $\sigma_Z$ component of the first qubit, see Eqs.~(\ref{eq: Tall}-\ref{eq: T1z}). The bottom line is the information of the label $I(T_{1}^Z:Y)$. The two bottom lines are the same as in the main figure. 
    }
    \label{fig:MI_plot}
\end{figure}

In Fig.~\ref{fig:MI_plot} we plot the quantities $I(X:T_{1}^Z)$ and $I(Y:T_{1}^Z)$ as functions of the learning epoch. As expected we see that the information on the data $I(X:T_{1}^Z)$ first jumps up and then gently comes down, which is a synonym for compression. In contrast, the information on the label $I(Y:T_{1}^Z)$ steadily goes up. In the inset of the figure Fig.~\ref{fig:MI_plot} we also compare $I(X:T_{1}^Z), I(X:T_{1})$ and $I(X:T_\text{all})$ to shine some light on how does the compression happens. As expected for unitary quantum circuits we see that a lot of compression happens when going from the global state $T_\text{all}\to T_{1}\to T_{z}^1$ to "the corner of the Hilbert space" which is directly used for classification.

\section{Compression-Gnostic learning}
\label{sec:problem_setting}

As we just have seen the information plane can help our understanding of the learning dynamics, also for quantum circuits. The key insight here is that by monitoring some correlations between the data and its internal representation inside the model, one can have an idea of how far the model is from a learned state (from a minimal sufficient statistics). Furthermore, this can be monitored during the learning process, and it is only natural to try to use this insight to improve the learning algorithm in the first place. We consider two different ways to improve the learning algorithm by monitoring the quantity $I(T:X)$ as a proxy for data compression. 
The first is to directly embed in the loss function. While the second is to embed it in the learning scheduler. In both cases, a key feature of our methods is that they can be used as add-ons to the existing algorithms.

\subsection{Compression-gnostic loss function}
\label{subsec:Compression-gnostic loss function}

Let us first focus on classification tasks. Our motivation here is to tweak the loss function to put pressure on the model to converge to minimal sufficient statistics since this is what we expect from its learned state. A simple way to do so is to consider the following loss 
$L_{\text{Comp}} := (I(T:X)-I_*)^2,$
which penalizes the deviation of $I(T:X)$ from its converged value $I_*$ that one would hope to obtain for a learned circuit.
Note that for a model that is close to a minimal sufficient statistics, we expect the inequalities $I(T:Y)\leq I(X:Y)$ and $I(T:Y)\leq I(Z:X)$ to be nearly saturated, as it should retain all the information pertinent to the label but nothing else. It is thus it is natural to set  $I_*=I(X:Y)$, which can be directly computed from the trained data. For datasets with continuous features the quantity is usually simply $I(X:Y)=H(Y)$
as all the data points $X_i$ are different. In general, for classification tasks setting $I_*= H(Y)$ and 
\begin{equation}
    L_{\text{Comp}}^C := (I(T:X)-H(Y))^2,
    \label{eq:new_loss_term}
\end{equation}
is a good rule of thumb.

For regression tasks, it is not straightforward to sensibly estimate what is the optimal value $I_*$ expected from a converged model. A simple solution is to remove this term from the loss function altogether, and simply penalize it whenever it retains more information on the data. This motivated the introduction of the following loss-term
\begin{equation}
    L_{\text{Comp}}^R := I(T:X).
    \label{eq:new_loss_term_reg}
\end{equation}

Finally, our motivation is to combine the new compression-based terms with the standard loss function $L_\text{Err}$ used in the algorithm we wish to improve. To do so we follow the approach of~\cite{peer2022improving}, where the authors introduced a regularization of the loss function to control the output entropy of different neural network layers. In a similar way, we define the regularized loss function as 
\begin{equation}\label{eq: novel loss}
    \mathcal{L} := L_{\text{Err}} \times (1+\alpha  \, L_{\text{Comp}}),
\end{equation}
where $\alpha$ is a new (dynamical) hyper-parameter of the model that controls the pressure from the mutual information term.
Combining the standard loss with $L_{\text{Comp}}$ this way ensures that minimizing the information is secondary and should not hurt the training. \\

\subsection{Compression-gnostic learning "scheduler"}

\label{subsec: scheduler}
Let us now focus on training based on the optimization of the loss function with the (stochastic) gradient descent method, commonly used in machine learning. Here the variational parameters of the model $\theta$ are updated based on the knowledge of the local loss landscape given by the gradient of the loss function $\nabla_\theta \mathcal{L}$ on the data (batch).

By the product rule for the modified loss function in Eq.~\eqref{eq: novel loss} the gradient decomposed as the sum of two terms 
\begin{equation}\label{eq: gradient full}
    \nabla_\theta \mathcal{L} = \left(\nabla_\theta L_{\text{Err}}  \right)\times  (1+\alpha L_{\text{Comp}}) + \alpha L_{\text{Err}} \times \left(\nabla_\theta L_{\text{Comp}} \right)
\end{equation}
that can be evaluated independently. Here, depending on how the data compression term $L_{\text{Comp}}$ or the mutual information $I(T:X)$ are estimated, the second term might not be straightforward to compute.
However, let us now assume that it is computed by applying the binning function
$\mathcal{B}$ in Eq.~\eqref{eq: disc map} to a deterministic continuous variable $T$ (given by $T=\mathcal{T}(\rho)$ for quantum models). As already pointed out in \cite{amjad2019learning}, in this case the mutual information $I(\mathcal{B}(T):X)$ is a piecewise constant function of the parameter $\theta$ (because $\mathcal{B}$ is a piecewise constant function of $T$). It follows that the gradient of the compression term $\nabla_\theta L_{\text{Comp}}=0$ vanishing almost everywhere on the parameter space, and trying to compute it is a waste of computational resources.

In this case, or whenever one wants to avoid computing  $\nabla_\theta L_{\text{Comp}}$, a simpler way to realize compression-based learning is to introduce it as a \textit{learning scheduler}. Concretely, we now only keep the first term in the gradient of the loss functions of Eq.~\eqref{eq: gradient full}
\begin{equation}\label{eq: gradient part}
    \nabla_\theta \mathcal{L} = \left(\nabla_\theta L_{\text{Err}}  \right)\times (1+\alpha L_{\text{Comp}}),
\end{equation}
such that the role of the compression term  $L_{\text{Comp}}$ is now not to modify the loss landscape, but to control the learning rate. The idea here is to refine the training process by adapting the learning rate based on the model's proximity to the learned state, a synonym of achieving optimal data compression. While the term $\nabla_\theta L_{\text{Loss}}$ tells the algorithm about the local features of the loss landscape  -- the gradient around its current state, the term $L_{\text{Comp}}$ tells it about an important global feature -- how far is the current model state from the optimum.

Let us illustrate it for a classification task. This scheduler operates by dynamically adjusting the magnitude of the gradient updates in accordance with the mutual information between the features and the target variable, $I(T:X)$, and the ideal mutual information value $I_*$. Specifically, when the mutual information $I(T:X)$ is significantly higher than $I_*$, indicating that the model retains more information than necessary and is far from a learned state, the learning scheduler amplifies the gradient updates.  This acceleration encourages rapid model adjustment towards minimizing unnecessary information retention, pushing the model closer to the minimal sufficient statistics for the task at hand. Conversely, as the model approaches optimal compression, where $I(T:X)$ nears $I_*$, the scheduler decreases the gradient magnitude, effectively reducing the learning rate. This fine-tuning phase ensures that the model converges smoothly to the optimal point without overshooting, thereby improving training efficiency and potentially leading to better generalization.

\subsection{The new hyperparamether $\alpha$}

The new loss function (or scheduler) we introduced contains a new parameter $\alpha$ which controls how strong the influence of the compression-based feedback on the learning. A priori a "good" value of $\alpha$ for a given learning task is not clear, hence we will treat it as an additional hyper-parameter of the algorithm. In particular, we will be interested in studying how different values of $\alpha$ (including the case with $\alpha=0$, i.e. no compression-based feedback) influence the results of training.

Generally, we expect that in the early learning epochs, the MI on data is growing rapidly (see Fig.~\ref{fig:MI_plot}). Using the terminology of \cite{shwartzziv2017opening} this can be viewed as the fitting phase of the learning, where the model learns to encode the data faithfully (followed by the later compression phase). This suggests that a high value of $\alpha$, penalizing high MI on data, might limit the learning process at the early stages. 
In order to avoid interfering with this phase of training, we also introduce a dynamic $\alpha$ given by
\begin{align}
    \alpha (\text{s}) =
    \begin{cases}
        (\alpha_\text{max}+1)^\frac{\text{s}}{s_\text{max}}-1
        & \text{if } \text{s}\le s_\text{max}\\
        \alpha_{\text{max}}, & \text{otherwise}
    \end{cases},
    \label{eq:dynamic_a}
\end{align}
that starts at $\alpha=0$ and slowly increases for the first few epochs before reaching its desired value (called $\alpha_{\text{max}}$) at the epoch $s_\text{max}$, where the fitting phase should be over. \\


\section{Results} \label{sec:results}

Now that we have introduced the novel learning algorithm based on information compression, we test its performance on several learning tasks. We primarily focus on the binary classification task tackled with parameterized quantum circuits (simulated on classical hardware). In the second step, we apply a similar model to a regression problem. Finally, we consider an example with classical neural networks.

\subsection{Architecture of the parametric quantum circuits.}

\begin{figure*}[!ht]
    \centering
    \includegraphics[width=\textwidth]{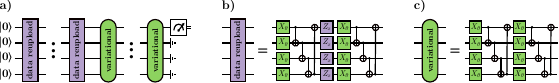}
    \caption{Architecture of the circuits for the quantum learning models for the case of $N=4$ qubits. \textbf{a)} The overall circuit is composed of several data reuploading layers in series followed by the variational layers. In the end, the first qubit is measured on a computational basis, and the expectation value of the corresponding Pauli-Z observable is the classical output by the circuit.  \textbf{b)} The internal structure of the variational layers, composed of parameter rotations $X_\theta$ around the Pauli-X, the rotations $Z_x$ around the Pauli-Z encoding the data features, and the two-qubit CNOT gates. The number of encoding gates $Z_x$ can vary depending on the dataset and the number of qubits in the circuit, that is, in a given data re-uploading layer the gates $Z_x$ are not necessarily present to act on each qubit. \textbf{c)} The inner structure of the variational layers, which do not directly depend on the input data.  }
    \label{fig:Full_circuit}
\end{figure*}

To start we briefly sketch the architecture of the quantum circuits, which will be considered later, illustrated in Fig.\ref{fig:Full_circuit}.\\

In all of the examples, the models are implemented with a $N$-qubit circuit (with $4\leq N\leq 12$) composed of two types of layers applied in series on the qubits all prepared in the ground state $\ket{0}$. The "data re-uploading layers" depend on both the features of the input data and variational parameters, these are followed by the "variation layers" which do not directly depend on the input data. The structure of both type of layers is outline in Fig.\ref{fig:Full_circuit}(b,c), they are composed of parametric single-qubit rotations  generated by $\sigma_X$ or $\sigma_Z$ and represented by the unitary operators 
\begin{equation} X_\theta= \exp(-\ii \frac{\theta}{2}
 \sigma_X) \qquad \text{or}\qquad Z_x= \exp(-\ii \frac{x}{2} \sigma_Z),
\end{equation}
and the nearest neighbor two-qubit CNOT gates (in the 1D topology with periodic boundary condition) arranged in a "ring". 

Note that each single qubit gate $X_\theta$ is specified by a different scalar variational parameter $\theta$, and the list of all variation parameters is denoted $\bm \theta$. Similarly, in the data reuploading layers a different scalar data feature $x^{(k)}$ is encoder on each qubit $k$ via the rotations $Z_x$ (depending on the number of qubits and the number of features the gates $Z_x$ might not be present on each qubit). In contrast, in the subsequent reuploading layers the same features are repeatedly encoded on the same qubits, hence the name of the layers.  The decision to include variational circuits between the feature encoding gates is motivated by this sequential structure. In this way, the data reuploading layers can already act as a pre-processing step, extracting significant features or representations of the data at the encoding. While the following variational layers can further process and compress the encoded information.\\

At the very end of the circuit only the first qubit is measured in the computational basis, the expected value $T_1^Z$ of the $\sigma_Z^{(1)}$ observable is output by the circuit and classically processed for the classification/regression task. The other qubits are discarded, playing the role of information sink. \\

In this work, we developed and simulated quantum circuits using the Pennylane library to implement a quantum neural network  \cite{bergholm2018pennylane}. The network architecture involves alternating between encoding classical data into quantum states using angle embedding and applying variational quantum layers with RX rotations. To optimize the quantum model, we employed various quantum-compatible optimizers such as Stochastic Gradient Descent \cite{NIPS2007_0d3180d6}. The simulations were conducted on the lightning.qubit backend, providing a controlled environment to fine-tune the quantum algorithms and assess their performance.


\subsection{Example 1: Implementation on synthetic dataset}
\label{sec:Implementation on synthetic dataset}

For the first example, we consider the binary classification task \cite{Yi_2021, Park_2020},
for a synthetic dataset (see Fig.~\ref{fig:3d_data} in the appendix). The dataset consists of $n=800$ points scattered in three-dimensional space $\bm x_i=(x_i^{(1)},x_i^{(2)},x_i^{(3)})\in \mathds{R}^3$. Despite its relatively small size, resulting in short training times, it is complex enough to contain irrelevant features that can be compressed during learning. 
The dataset consists of four "overlapping" clouds of 200 points each, with two clouds representing the class 0 (labeled with $Y=-1$) and the other two representing the class 1 (labeled with $Y=1$), implying $H(Y)=1$. Each cloud is sampled from a multivariate normal distribution\\

The parametric quantum circuit we use is structured as depicted in Fig.~\ref{fig:Full_circuit}, with four qubits, three data reuploading, and two variational layers. Given that the data set is three-dimensional, the features are only encoded in the three first qubits. The estimator for the label is taken to be the sign of the expected value of $\sigma_Z$ observable on the first qubit
\begin{equation}
 y_i = \text{sign}\,  T_{1}^Z = \text{sign} \left(\tr \rho_{Q_1\dots Q_4} \, \sigma_Z^{(1)}\right)
\end{equation}
in depends on the input $\bm x_i$ and the parameters $\bm \theta$ via the state $\rho_{Q_1\dots Q_4}(\bm x_i,\bm \theta) $ prepared by the circuit.\\

To compute the mutual information term $I(T:X)$ we discretize the value $t=T_1^Z$ down to $m=6$ equals intervals between -1 and 1. Each data point $\bm x_i$ is thus mapped into a value $z_i =1,\dots m$, depending on which interval the value  $T_1^Z$ falls in. We then compute the mutual information on the data $I(Z:X)=H(Z)+H(X)-H(Z,X)$.
Note that in our case all the data points are different and the encoding $X\to Z$ is deterministic. Hence $H(Z,X)=H(X)=\log(n)$, and the mutual information is simply given by the entropy of $Z$
\begin{equation}
I(Z:X) = H(Z) = - \sum_{k=1}^m \text{Pr}(z_i=k) \log \text{Pr}(z_i=k)
\end{equation}
where $\text{Pr}(z_i=k)$ is the total number of data points that fall in the interval $k$ divided by $n$. This quantity has a nontrivial behavior as several data points are mapped into the same value $z_i$. Finally, for our dataset $I_* = I(X:Y)=H(Y)=1$, and we have now introduced all the elements required to compute the loss function $L_{\text{Comp}}^C$ in Eq.~\eqref{eq:new_loss_term}.
    
 Here we explained how we estimate the information on the data $I(T_1^Z:X)$  (or the label $I(T_1^Z:Y)$) carried by the expected value of the $\sigma_Z$ observable on the first qubit. In a similar way, we can compute the information on data carried by the full state of the first qubit $I(T_1:X)$ and the state of all the qubits $I(T_\text{all}:X)$. When working with continuous variables the mutual information and entropy computation need to be adapted,  as detailed in the appendix~\ref{sec: Mutual information for continuous variables} where we discuss the discretization process for quantum states. For $\alpha=0$, i.e. the MSE loss function, the evolution of these quantities during learning is depicted in Fig.~\ref{fig:MI_plot}. \\
 
\begin{figure*}[t!!]
    \centering
    \includegraphics[width=170mm,scale=0.55]{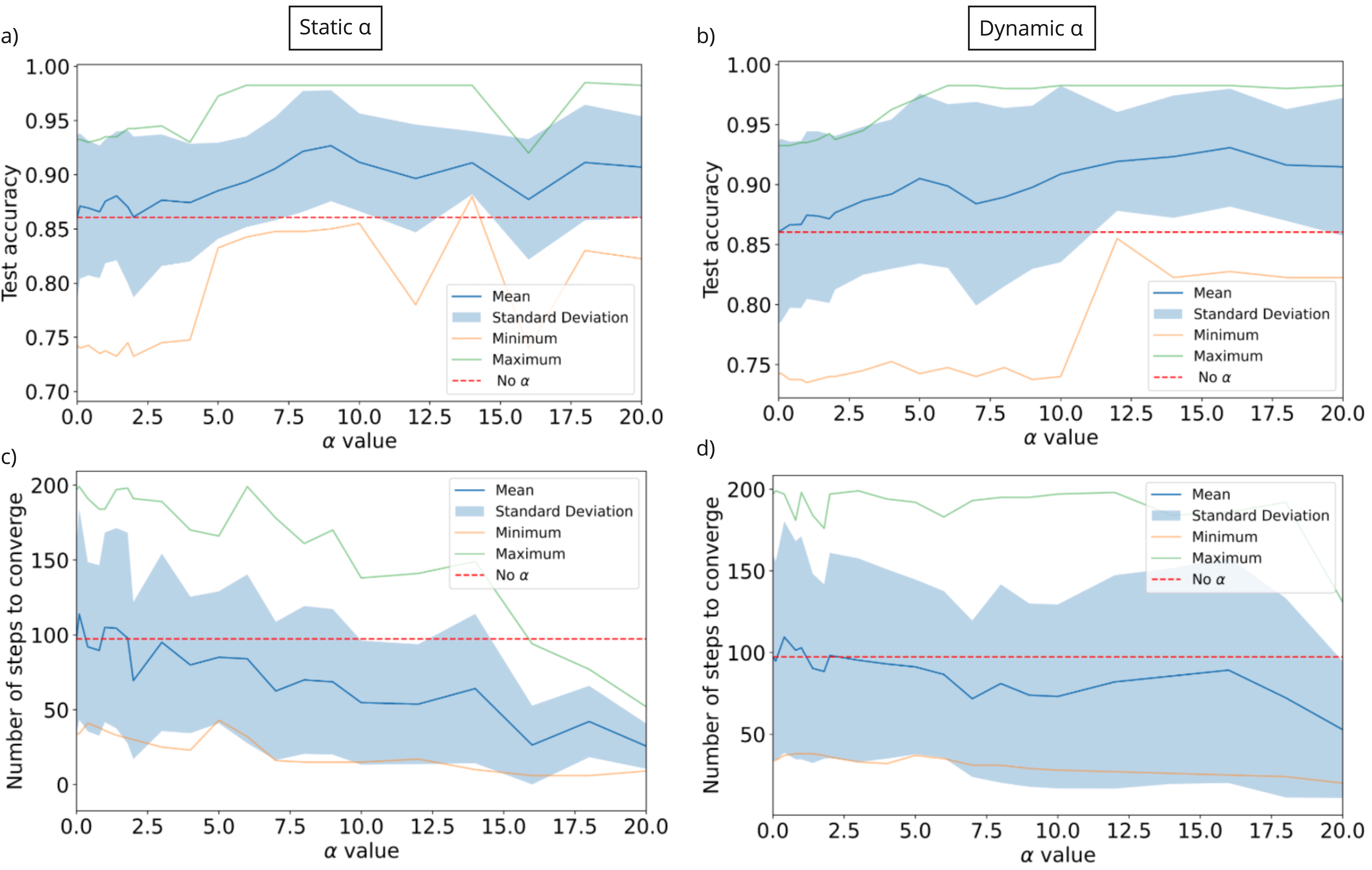}
    \caption{In a) and b) plot of the test accuracy obtained on the test set compared to the $\alpha$ values for static and dynamic $\alpha$. The blue curve and background show mean values and start deviation. The red dashed line shows the models with $\alpha = 0$. Maximum and minimum show the best and worst models. In c) and d) plot of the number of steps to converge compared to the $\alpha$ values for static and dynamic $\alpha$.  The blue curve and background show mean values and start deviation. The red dash line shows the models with $\alpha = 0$. Maximum and minimum show the best models within various initial weights.}
    \label{fig:Results1}
\end{figure*}

\begin{figure*}[t!]
    \centering
    \includegraphics[width=170mm,scale=0.55]{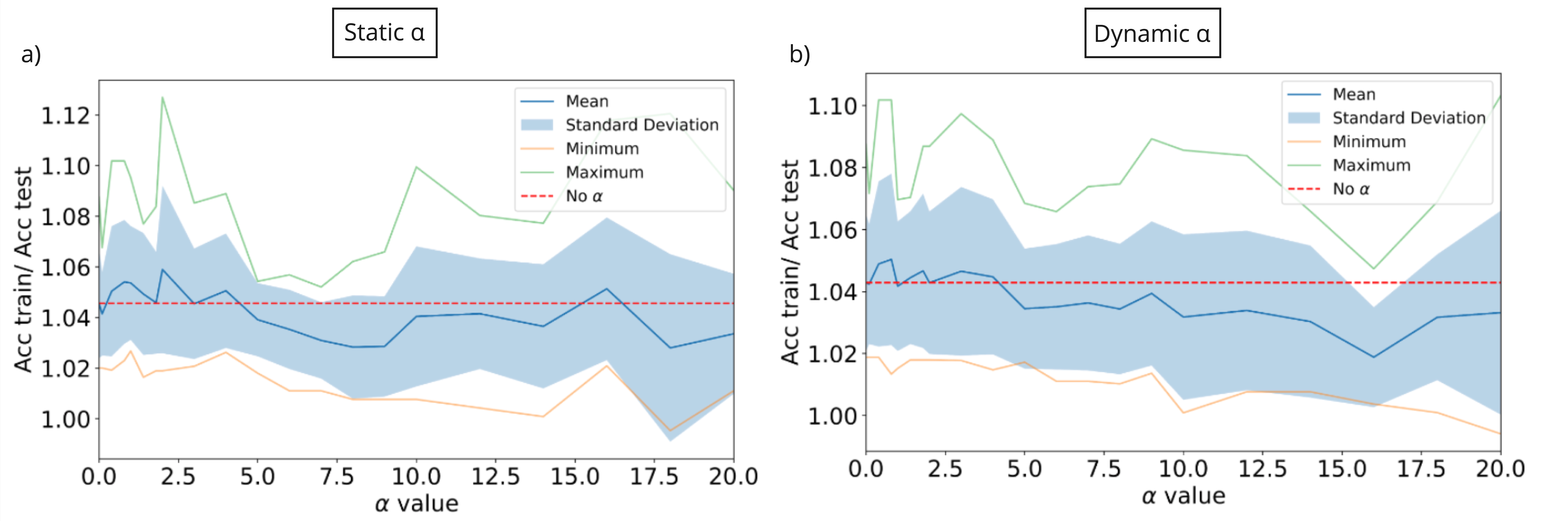}
    \caption{In a) and b) plot of the training and test accuracy ratio for static and dynamic $\alpha$ values. The blue curve and background show mean values and start deviation. The red dash line shows the models with $\alpha = 0$. Maximum and minimum show the best models within various initial weights.}
    \label{fig:Results2}
\end{figure*}

 Next, we discuss how turning on the compression-based feedback affects the performance of the learning algorithm. The central question is whether introducing this novel loss function contributes to increased model performances. To investigate this, we consider several metrics to benchmark the performance of the model: the mean test accuracy Fig.~\ref{fig:Results1}(a and b), the number of steps to converge Fig.~\ref{fig:Results1}(c and d), or the capability to generalize to unseen data Fig.~\ref{fig:Results2}. 

To understand the importance of the new term in the loss function in Eq.~\eqref{eq:new_loss_term} and its overall influence on the model, we systematically vary the $\alpha$ parameter and compare the performance of the resulting models.  Considering the limited size of the dataset, the choice of initial weights can highly influence the performance of the model. 
To soften any potential advantages stemming from a lucky selection of initial weights (which might benefit certain $\alpha$ values by pure luck), for each value of $\alpha$ the models are executed using a total of 10 distinct initial weight configurations. To define each configuration all the variational parameters are sampled from a uniform distribution on an interval $[0,1]$.\\

\subsubsection{Test Accuracy}

First, let us analyze the accuracy achieved on the test set for different values of $\alpha$, as illustrated in Fig.~\ref{fig:Results1}(a) for static $\alpha$ and Fig.~\ref{fig:Results1}(b) for dynamic $\alpha$.   
The value $\alpha = 0$ corresponds to ignoring the mutual information term in Eq.~\eqref{eq:new_loss_term} and corresponds to the standard MSE loss function. This defines the baseline for the comparison, given by the red dashed line in Figs.~\ref{fig:Results1}. 
The blue line and the light blue region depict the mean and standard deviation of the test accuracy (over different initializations).

By analyzing the mean for static $\alpha$ (Fig.\ref{fig:Results1}a)  we conclude that the accuracy improves as $\alpha$ increases, going from 86\% mean test accuracy for $\alpha=0$ up to 93\% mean test accuracy. It is worth noting that all models with positive $\alpha$ values surpass the accuracy achieved by the models with no $\alpha$, indicating that incorporating a non-zero $\alpha$ consistently enhances the test accuracy.\\

For dynamic $\alpha(s)$ results (Fig.~\ref{fig:Results1}b)  we used the dependence on the learning step given in Eq.~\eqref{eq:dynamic_a} for $s_\text{max}=30$ and varying $\alpha_\text{max}$ corresponding to the x-axis ($\alpha$ value) in the plots. When comparing these results with the case of static $\alpha$, we observe a more linear and consistent increase in mean test accuracy. The results indicate that incorporating a dynamic $\alpha$ leads to improved stability of the models and higher overall test accuracy. Moreover, for high values of $\alpha$, the test accuracy achieved by the worst models (yellow curve in Fig.~\ref{fig:Results1}b) is comparable to the mean test accuracy of the models with no $\alpha$, demonstrating significant test accuracy improvement. \\

\subsubsection{Model convergence speed}

Second, we examine whether incorporating a mutual information term results in faster convergence of the models, which is to be expected since the novel term pressures the model to converge to a minimal sufficient statistics.
For static $\alpha$, Fig.~\ref{fig:Results1}(c) illustrates the number of steps (i.e. number of epochs) required for the models to converge to their maximum test accuracy. Notably, incorporating a non-zero $\alpha$ not only decreases the number of steps taken to converge but also significantly reduces its standard deviation. With the MSE loss ($\alpha=0$), the models need an average of 100 steps to converge. With high $\alpha$, this number drops to less than 50. \\

As for the dynamic $\alpha$ case, depicted in Fig.~\ref{fig:Results1}d, we see that it has a less significant impact on the convergence rate as compared to the static case. The average number of steps required to converge remains relatively close to the 100-step baseline across a wide range of $\alpha$ values. The standard deviation is also particularly high. These results are however also expected since introducing a slow onset of the $\alpha$ parameter delays the influence of the mutual information term which is responsible for the speedup of the model convergence.\\

It is worth noting that when running models with $\alpha$  (or $\alpha_\text{max}$) values exceeding 20, the observed advantage does not appear to increase further, and the standard deviation begins to rise (see Appendix,  Fig.~\ref{fig:high_a_Steps}).

\subsubsection{Generalization}

Finally, we examine the generalization capability of the models with the novel loss function. To have a grasp of it we consider the ratio of the training and test accuracies. These quantities represent the ability of the model to correctly classify the data on the training versus unseen data. A lower ratio is thus a synonym of better generalization. \\

We found that the models with static $\alpha$ did not display a conclusive improvement in this metric, as shown in  Fig.~\ref{fig:Results2}a. In contrast, models with high dynamic $\alpha$ seem to improve generalization. In Fig.~\ref{fig:Results2}b we observe a stable decrease in the ratio with increasing $\alpha$ values. \\

The usual plausible explanation of this effect is that the regularization, coming form the addition of the compression term,  makes the loss landscape sharper and leads to an effective reduction of the expressivity of the model (attainable during learning). The general logict of statistical learning theory then suggests that a less expressive model is less prone to over-fitting, and thus generalizes better.

\subsubsection{Summary} \label{sec:discussion}

We have seen that both approaches, static $\alpha$ and dynamic $\alpha$, improve the performance of the model as compared to the standard case $\alpha =0$. The results are summarized in the following table.
\begin{table}[h]
\centering
\small 
\begin{tabular}{|p{2.5cm}|p{2.5cm}|p{2.5cm}|p{2cm}|} 
\hline
& \text{Static $\alpha$} & \text{Dynamic $\alpha$}  \\
\hline 
\text{Test accuracy} & Increase of $\mathbf{7\%}$ & More stable models \\
\hline 
Number of steps to converge & Divided by $\mathbf{2}$ & Divided by $1.3$  \\
\hline 
Test accuracy after 25 steps & Increase of $13\%$ & Increase of $5\%$  \\
\hline 
Helps for generalization & Inconclusive & Yes  \\
\hline
\end{tabular}
\end{table}\\

In general, employing a static $\alpha$ leads to an increase in the mean test accuracy of 7\% while at the same time significantly reducing the number of steps required to converge (by a factor of two in the best case).
On the other hand, utilizing a dynamic $\alpha$ results in more stable models with respect to mean test accuracy and slightly increases the models' generalization capabilities. A tempting explanation for these enhancements is that the novel loss function helps the model to compress better, i.e. reduce the  MI on data, as illustrated in Fig.~\ref{fig:MI_compression} of the Appendix.
Nevertheless, in both cases (static and dynamic), running with $\alpha$ values that are too high eventually starts to deteriorate the model's performance, meaning that the optimal value has to be chosen carefully.

\subsection{Other examples}

\subsubsection{Real world classification examples}

This study evaluates the performance of quantum models on two real-world datasets: the California Housing Price Prediction dataset \cite{noauthor_california_nodate} and the Stroke Prediction dataset  \cite{noauthor_heart_nodate, noauthor_stroke_nodate}, more details are attached in Appendix \ref{sec: california dataset} and Appendix \ref{sec: Stroke Prediction Dataset}.

The California Housing dataset includes features such as Median Income and House Age, used to predict whether house prices are above or below the median. A similar architecture used in Section \ref{sec:Implementation on synthetic dataset} of the quantum circuit was used. It contains 4 qubits and 5 layers were utilized. We focused on the impact of the hyperparameter $\alpha$. The results showed that with $\alpha = 15$, the model achieved a test accuracy of 0.812 and converged in 104 steps. In contrast, without $\alpha$ (i.e., $\alpha = 0$), the model had a lower test accuracy of 0.770 and required 174 steps to converge. Any non-zero value of $\alpha$ consistently improved both accuracy and convergence speed.

For the Stroke Prediction dataset, which predicts the likelihood of a stroke based on health-related features, a quantum circuit with 12 qubits and 3 layers was employed due to the higher number of features. The model’s performance was measured using the Area Under the Curve (AUC) due to the dataset’s imbalance. The introduction of $\alpha$ led to a significant improvement, increasing the AUC from 0.68 to 0.73 on the test set. Models with non-zero $\alpha$ values consistently showed better performance.

Across both datasets, the inclusion of the $\alpha$ parameter consistently enhanced model accuracy and efficiency, reducing the number of steps needed for convergence. These findings highlight the potential of $\alpha$ to optimize quantum models in practical applications.
However, while these initial results are encouraging, it is important to note that they would require further confirmation through more extensive testing. A comprehensive analysis, including varied initial seeds and additional datasets, is necessary to fully validate the robustness and generalization of these findings.

\subsubsection{Regression example}

We explored applying our training method to regression tasks using a dataset from a photovoltaic (PV) power generation plant in the Mediterranean region \cite{MALVONI20161639}, more details are provided in Appendix \ref{sec: Regression example}. 
The dataset spans 21 months, providing hourly measurements of PV power output and meteorological variables like ambient temperature and solar irradiance. 
After adopting the loss-term for regression, our objective was to enhance the results achieved in \cite{sagingalieva2023photovoltaic} by introducing a non-zero $\alpha$ value.

The model employed is a hybrid of classical and quantum components. 
The classical part consists of fully connected neural network layers, starting with an input layer of 42 neurons, followed by ReLU activations, and reducing dimensionality to match three qubits for the quantum component. 
The quantum part utilizes a 3-qubit Variational Quantum Circuit (VQC) with three layers, featuring strong entangling layers via CNOT gates and optimized rotations around the Z-axis. 
After processing, the quantum output is fed into a final layer to predict PV power.

The model was trained with a learning rate of 0.005 over 10 epochs, and performance was evaluated using the R-squared ($\text{R}^2$) score.
As shown in Figure \ref{fig:Results_pvpower}, the introduction of non-zero $\alpha$ values generally improves the $\text{R}^2$ scores compared to the baseline of $\alpha = 0$.
Peaks in performance were observed at $\alpha = 5$ and $\alpha = 18$, indicating that non-zero $\alpha$ values enhance the model’s ability to capture underlying data patterns, extending the benefits seen in classification tasks.

\subsubsection{Classical neural network example}

\label{subsec: classical}
Finally, we investigated whether the concept of regularization through the parameter $\alpha$ could be applied to classical neural networks using a water potability dataset \cite{noauthor_water_nodate}.  more details are contained in Appendix \ref{sec: Example with neural network model}.
The dataset includes features such as pH, Hardness, and Solids, with the target variable being a binary classification label indicating water potability.

The neural network architecture consisted of three fully connected dense layers: 
the input layer with 9 features, followed by two hidden layers with 36 neurons each, both including 50\% dropout to prevent overfitting. 
ReLU activation functions were applied to the hidden layers, and a sigmoid function was used in the output layer for binary classification.

We compared models with $\alpha = 0$ (baseline) and non-zero $\alpha$ values, testing 20 different learning rates ranging from 0.0001 to 0.4 for each $\alpha$ value. Early stopping was implemented to halt training if no improvement in validation accuracy was observed for 50 epochs. The results averaged across multiple runs, revealed that only the model with $\alpha = 20$ showed a slight improvement in test accuracy, increasing from approximately 66\% to 67\%. Other non-zero $\alpha$ models demonstrated lower accuracy compared to the baseline.

Additionally, as shown in Figure \ref{fig:Results_MI_classical}, models with non-zero $\alpha$ values generally converged faster, suggesting that while regularization may speed up training, it does not necessarily improve accuracy. These findings are preliminary, and further analysis with a broader range of $\alpha$ values and additional metrics is needed to draw definitive conclusions about the impact of regularization on model performance.

\section{Discussion}
\label{sec:conclusion}

One expects a trained machine learning model to be able to distill compressed representations of the input data, which retain features pertinent for the training task but discard irrelevant information. Intuitively, the level of compression that a model achieves can be quantified by the mutual information $I(T:X)$ between the input data $X$ and its latent representation $T$ inside the model. Yet, when these variables are continuous-valued and the model  deterministic the bare mutual information completely misses this purpose. Hence, alternative ways to quantify the compression achieved by classical learning models have been proposed in the literature, e.g.~\cite{alemi2016deep,chalk2016relevant,shwartzziv2017opening,kolchinsky2019nonlinear,amjad2019learning,cvitkovic2019minimal}.\\

In this paper, we studied the same question of data compression in the context quantum machine learning models. Concretely, in Sec.~\ref{sec:inf QML} we proposed a way to estimate the compression of data achieved at different stages of a parametrized quantum circuit, inspired by the idea of discretization of continuous variables used in \cite{shwartzziv2017opening} to study the learning dynamics of deep neural networks.\\

In a second step (Sec.~\ref{sec:problem_setting}), we studied how a given learning algorithm can be improved by monitoring the level of compression attained by the current state of the model, and using this information when updating this state. Concretely, we proposed two ways to implement such compression-gnostic feedback. First (Sec.~\ref{subsec:Compression-gnostic loss function}), it can be realized by regularizing the loss function with a multiplicative term depending on the compression level. This is similar to the implementation of the entropy-batch regularization studied in \cite{peer2022improving}.
Second (Sec.~\ref{subsec: scheduler}), for (stochastic) gradient descent based algorithms the compression-gnostic feedback can be realize as a "scheduler", controlling the learning rate based on the level of compression achieved by the current state of the model. This approach does not modify the loss landscape, but also does not requires to differentiate the compression term. As we note, albeit different in general, in our implementation both approaches turn out to be equivalent. This is because the discretization based~\cite{shwartzziv2017opening} approach to estimate compression leads to a piecewise constant function (without gradient), as already noted in \cite{amjad2019learning}. Hence, we do not compare the two approaches.\\

To benchmark our ideas we probe our algorithm on several supervised learning tasks, both for classification and regression (Sec.~\ref{sec:results}). In most of our experiments  we simulated a quantum machine learning model on classical hardware. The obtained results are quite promising, as the introduction of the compression based feedback systematically improved the accuracy and convergence speed of the model. We also performed a preliminary study (Sec.~\ref{subsec: classical}) of the how such modification of the algorithm affects the performance of classical neural network on a binary classification task. 

\section*{Acknowledgements}

MK's contributions to this work were made while affiliated with Terra Quantum AG.

\bibliography{arxiv.bib}

\appendix

\label{sec: appendix}

\section{Additional information on the studied examples.}

\subsection{Additional figures for the synthetic dataset example}

\begin{figure}[h!]
    \centering
    \includegraphics[width=\columnwidth]{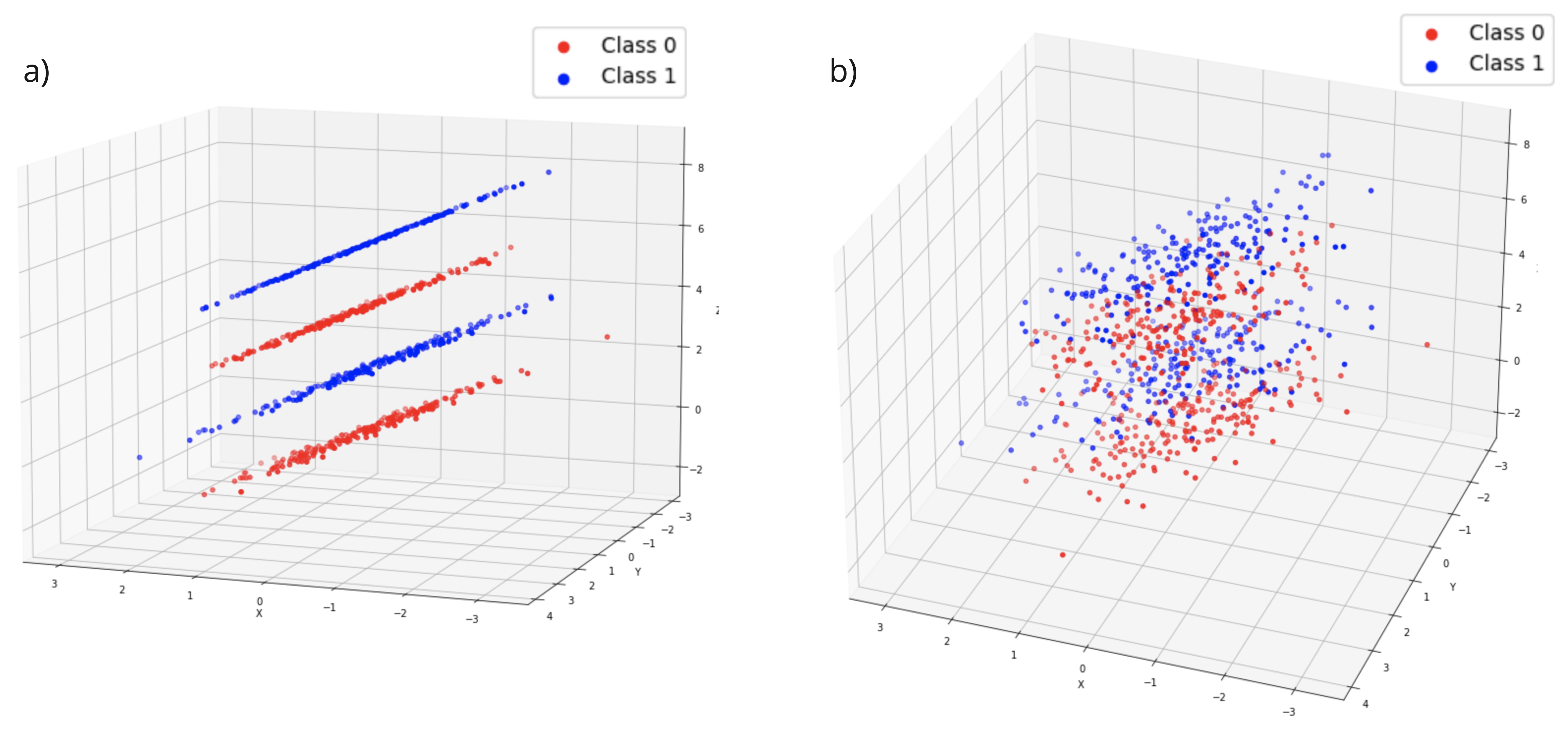}
    \caption{Visualization of the dataset from different angles. In $a)$, the four could, inclined by an angle, can we well distinguish. In $b)$, the normal distribution of each cloud in the dataset can be pictured. The points of color red belong to Class 0, while the points of color blue correspond to Class 1. The clouds are centered around the point (0, 0) in the $x-y$ plane. The $z$ coordinates values for the clouds are set according to the equation $z = a + x\cos(60)$, where $x$ represents the $x$ coordinates values of the points and $a$ take value in the interval $[0,2,4,6] $ for each of the four clouds. }
    \label{fig:3d_data}
\end{figure}

\begin{figure}[h!]
    \centering
    \includegraphics[width=\columnwidth]{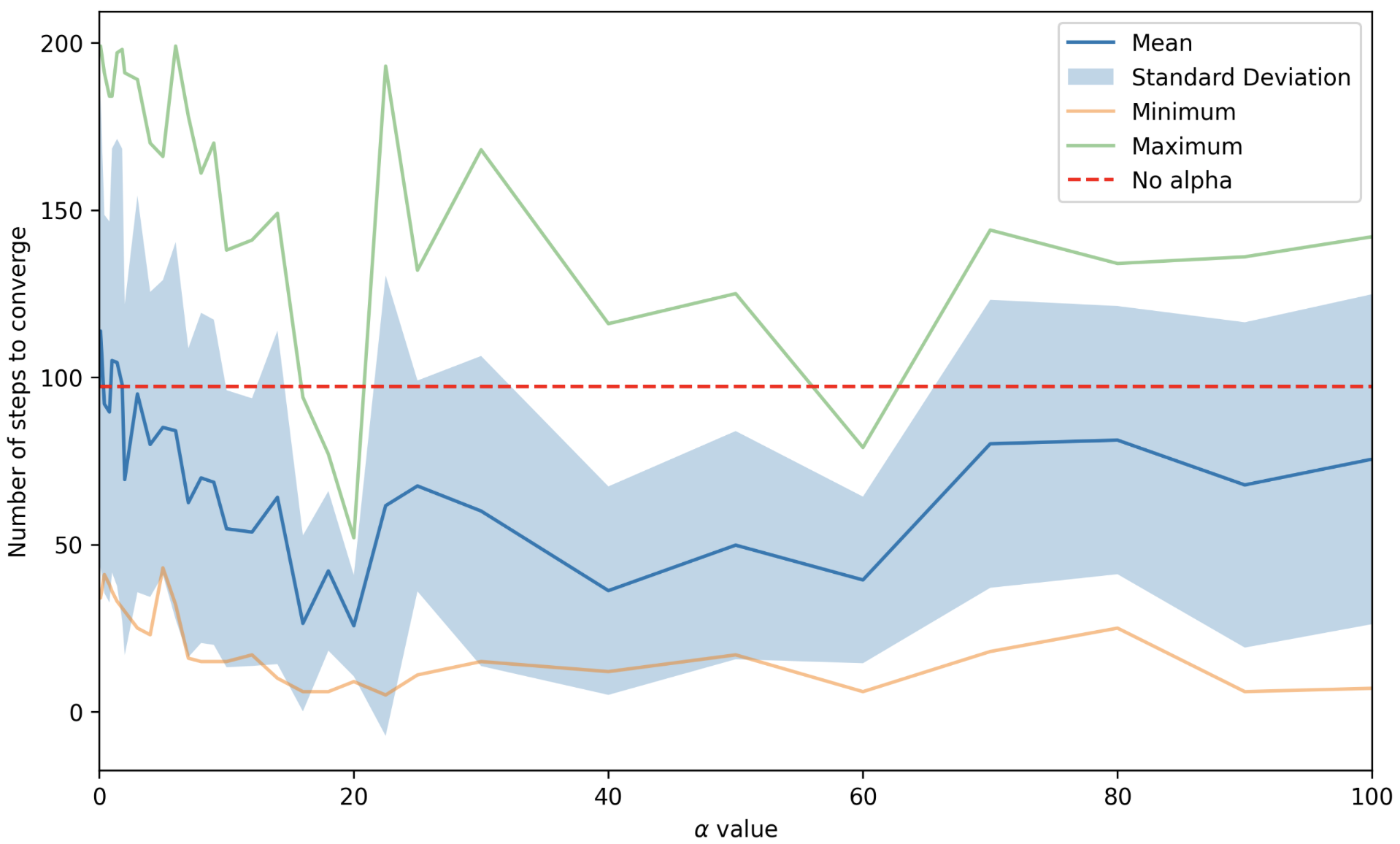}
    \caption{This figure depicts the relationship between the $\alpha$ value and the number of steps required for convergence. The results reveal that for $\alpha >20$, the number of steps to converge tends to increase, accompanied by a corresponding increase in the standard deviation.}
    \label{fig:high_a_Steps}
\end{figure}

\begin{figure}[h]
    \centering
    \includegraphics[width=\columnwidth]{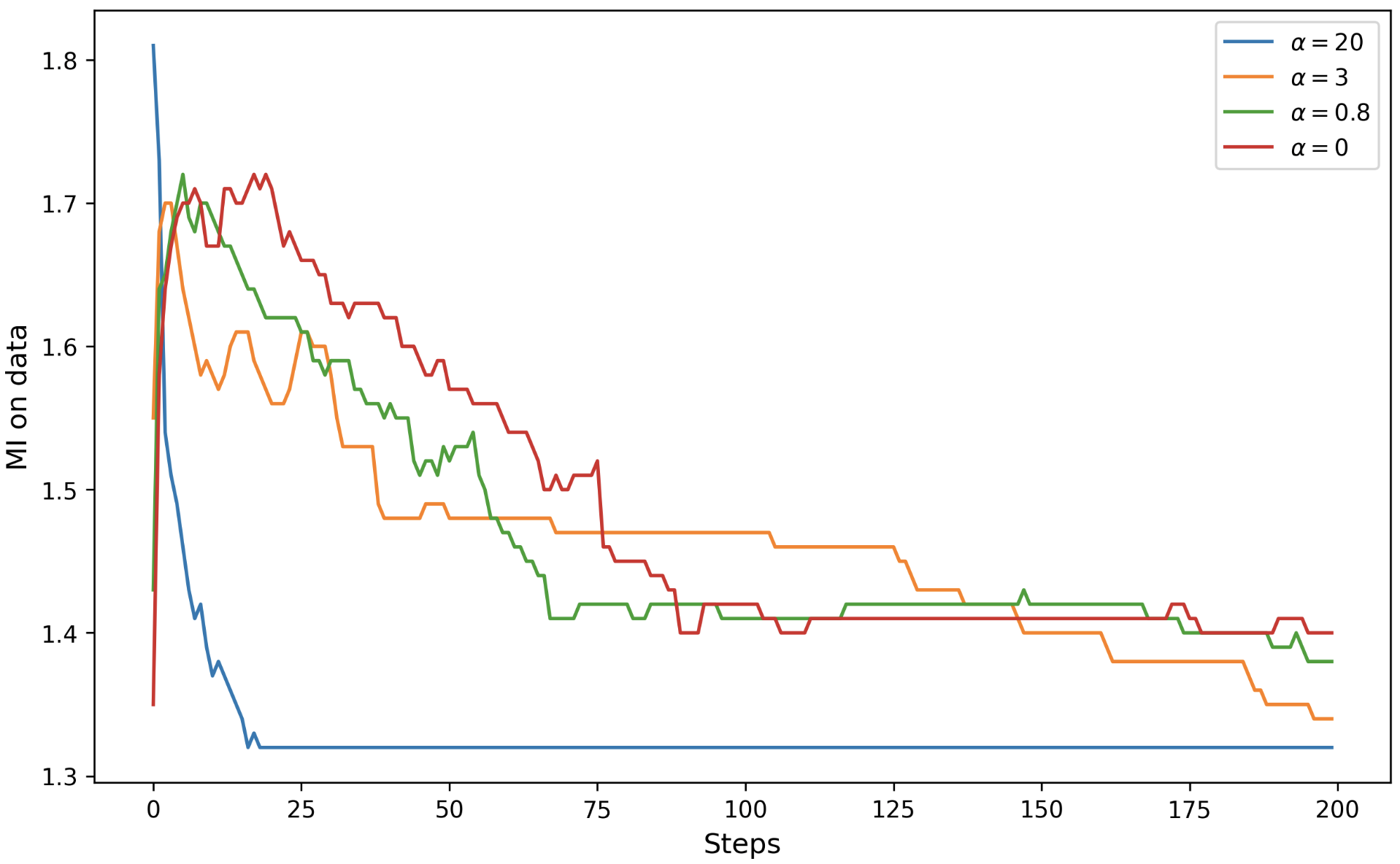}
    \caption{This figure illustrates the relationship between MI on data and the number of steps in the training process. It includes four curves representing different $\alpha$ values. High $\alpha$ values correspond to superior compression abilities.}
    \label{fig:MI_compression}
\end{figure}

The Figures \ref{fig:3d_data}-\ref{fig:MI_compression} mentioned in the main text give additional information on the synthetic dataset example.

\newpage

\subsection{California dataset}
\label{sec: california dataset}

The California Housing Price Prediction dataset \cite{noauthor_california_nodate} is a comprehensive collection of housing-related features designed for the task of predicting whether the price of a house is above or below the median price in various neighborhoods in California. This dataset is particularly valuable for machine learning and data analysis projects focused on real estate market analysis and housing price prediction.\\
The dataset is composed of eight features that characterize the houses:
\begin{itemize}
\item \text{Median Income:} Represents the median income of residents in a given neighborhood, providing insights into the income levels of the local population.
\item \text{House Age:} An integer value indicating the average age of houses in the neighborhood, serving as an indicator of the overall condition and age of the housing stock.
\item \text{Average Rooms:} A floating-point value representing the average number of rooms per dwelling in a neighborhood, offering insights into typical house sizes.
\item \text{Average Bedrooms:} Represents the average number of bedrooms per dwelling in the neighborhood, providing information about the housing configurations in the area.
\item \text{Population:} A floating-point value indicating the total population of a neighborhood, aiding in understanding the population density of an area.
\item \text{Average Occupancy:} The average occupancy per dwelling in a neighborhood, indicating how many people, on average, occupy each dwelling unit.
\item \text{Latitude:} A geographic coordinate specifying the north-south position of a location, helping to pinpoint the geographical location of neighborhoods in California.
\item \text{Longitude:} A geographic coordinate specifying the east-west position of a location, working in conjunction with latitude to precisely locate neighborhoods.
\end{itemize}

The first part of the work was on data preparation. The dataset was loaded from a text file and converted into a pandas DataFrame. We checked for missing values, but none were found. A boxplot was used to visualize the distribution of the features, which helped in identifying any potential outliers. To ensure that the features were on a comparable scale, we applied the \texttt{MinMaxScaler} to normalize all feature values between 0 and 1.
Because of the high number of features we opt for dimensionality reduction. To streamline the analysis and reduce computational complexity, we performed Principal Component Analysis (PCA) to reduce the dataset to three principal components. This step helps to retain the most significant variance in the data while simplifying the model’s input space.
The training process for the California Housing Price Prediction model was carried out using a similar QNN used in the synthetic dataset analysis, Section \ref{sec:Implementation on synthetic dataset} (which includes 4 qubits and 5 layers). The training data was split into training, validation, and test sets. The training process involved 200 epochs, with a batch size of 10, and a learning rate of 0.05. To optimize the model, we used the Adam optimizer. The inclusion of the $\alpha$ parameter was systematically varied across different runs to evaluate its impact on the model’s convergence speed and accuracy.
The table below clearly illustrates that the incorporation of $\alpha$ systematically enhances the model’s ability to achieve higher accuracy faster.

\begin{table}[h]
\centering
\small 
\begin{tabular}{|p{1cm}|p{3cm}|p{4cm}|p{2cm}|} 
\hline
& \text{Test accuracy} & \text{Steps to converge}  \\
\hline 
$\alpha$ = 0 & 0.77 &174 \\
\hline 
 $\alpha$ = 2 & 0.785 & 53  \\
\hline 
 $\alpha$ = 5 & 0.786 & 180  \\
\hline 
 $\alpha$ = 10 & 0.811 & 121  \\
\hline
 $\alpha$ = 15 & 0.812 & 104  \\
\hline
 $\alpha$ = 20 & 0.811 & 103  \\
\hline
\end{tabular}
\end{table}

\subsection{Stroke Prediction Dataset}
\label{sec: Stroke Prediction Dataset}
The Stroke Prediction Dataset \cite{noauthor_heart_nodate, noauthor_stroke_nodate} is designed to predict the likelihood of an individual experiencing a stroke based on various health-related attributes. This dataset is crucial in the context of healthcare, as strokes are among the leading causes of death globally, according to the World Health Organization (WHO).
The dataset includes a range of attributes such as:
\begin{itemize}
\item \text{ID:} A unique identifier for each entry.
\item \text{Gender:} Categorical values (“Male,” “Female,” “Other”).
\item \text{Age:} Continuous variable denoting the patient’s age.
\item \text{Hypertension:} Binary indicator (0 for no hypertension, 1 for hypertension).
\item \text{Heart Disease:} Binary indicator (0 for no heart disease, 1 for heart disease).
\item \text{Ever Married:} Categorical values (“No,” “Yes”).
\item \text{Work Type:} Categorical values (“Children,” “Govt job,” “Never worked,” “Private,” “Self-employed”).
\item \text{Residence Type:} Categorical values (“Rural,” “Urban”).
\item \text{Avg Glucose Level:} Continuous value representing the patient’s average blood glucose level.
\item \text{BMI:} Continuous variable indicating the body mass index.
\item \text{Smoking Status:} Categorical values (“Formerly smoked,” “Never smoked,” “Smokes,” “Unknown”).
\item \text{Stroke:} Binary target variable (1 if the patient has had a stroke, 0 otherwise).
\end{itemize}

During data preprocessing, missing values were identified and subsequently handled to ensure the integrity of the dataset. Each feature was then normalized using the \texttt{MinMaxScaler}, ensuring that the model could process the data efficiently. Categorical variables were encoded into numerical features to facilitate their use in the model. For example, the “gender” attribute was converted into binary or one-hot encoded representations. The final dataset, prepared for binary classification, consisted of 24 features, which included both the original and the newly generated features from categorical attributes.
To optimize the dataset for quantum computing applications, Principal Component Analysis (PCA) was employed, reducing the feature set from 24 to 12 principal components. This reduction not only decreased the computational load but also minimized the number of qubits required, making the model more efficient for quantum processing.
The model for stroke prediction was built using a QNN with 12 qubits and 3 layers. The training dataset was split into training, validation, and test sets. The model was trained over 200 epochs with a batch size of 100, using a learning rate of 0.1. 
To fine-tune the model, we employed the Adam optimizer. 
The $\alpha$ parameter was adjusted systematically across different runs to assess its effect on the model's convergence speed and accuracy. 
The table below demonstrates that incorporating $\alpha$ consistently improves the model's ability to achieve higher accuracy more quickly.

\begin{table}[h!]
\centering
\small 
\begin{tabular}{|p{1cm}|p{3cm}|p{4cm}|p{2cm}|} 
\hline
& \text{AUC test}  &\text{Steps to converge}  \\
\hline 
 $\alpha$ = 0 & 0.688 &184 \\
\hline 
 $\alpha$ = 2 & 0.735 &80 \\
\hline 
 $\alpha$ = 5 & 0.698  & 55\\
\hline 
 $\alpha$ = 10 & 0.701  & 75\\
\hline
 $\alpha$ = 15 & 0.688  &63 \\
\hline
 $\alpha$ = 20 & 0.696 & 45 \\
\hline
\end{tabular}
\end{table}

\subsection{Regression example}
\label{sec: Regression example}
The next question we encountered was how to apply this training method not only to classification tasks but also popularly solve in machine learning regression tasks. After an adaption of the loss-term described in Section~\ref{subsec:Compression-gnostic loss function} we are left with Eq.~\ref{eq:new_loss_term_reg} our new loss-term for the regression task.
The dataset used for this task is a publicly accessible dataset from a conventional photovoltaic power (PV) generation plant located in the Mediterranean region \cite{MALVONI20161639}. It spans a period of 21 months, providing hourly measurements of mean PV power output along with meteorological variables such as ambient temperature, module temperature, and solar irradiance. The dataset is comprehensive, covering over 500 days of data, and has been meticulously preprocessed to ensure accuracy, including the correction of anomalies and the handling of missing data.
The objective is to enhance the results achieved in \cite{sagingalieva2023photovoltaic} by introducing a non-zero $\alpha$ value. To accomplish this, we employ a hybrid model that combines classical and quantum components, following the architecture detailed in the mentioned study, which provides a comprehensive description of the model's design.
For clarity, let’s briefly describe the classical and quantum components of this hybrid model.
The classical component of the model consists of a series of fully connected neural network layers. Initially, the input data, which includes multiple features across a time window, is flattened to gather all features into a single vector. The first fully connected layer comprises 42 neurons and is followed by a ReLU activation function to introduce non-linearity. The output of this layer is passed to a second fully connected layer, which reduces the dimensionality to match the number of qubits, in this case, three, followed by another ReLU activation.

The quantum component involves a 3-qubit Variational Quantum Circuit (VQC) with three layers. The circuit employs “strong” entangling layers, implemented via CNOT gates, allowing for complex correlations between the qubits. Rotations are applied around the Z-axis, which are parameterized and optimized during the training phase. After quantum processing, the qubits are measured in the Y-basis. The quantum layer’s output is then passed through a final fully connected layer to produce the forecasted PV power output.

The model was trained with a learning rate of 0.005 over 10 epochs, using a batch size of 128. To evaluate the model’s performance, the R-squared ($\text{R}^2$) score was employed. The $\text{R}^2$ score is a statistical measure that indicates how well the model’s predictions align with the observed data, specifically quantifying the proportion of variance in the dependent variable that can be explained by the independent variables. A higher $\text{R}^2$ value signifies a model that more accurately captures the underlying patterns in the data, with a score of 1 indicating a perfect fit.

\begin{figure}[ht]
    \centering
    \includegraphics[width=\columnwidth]{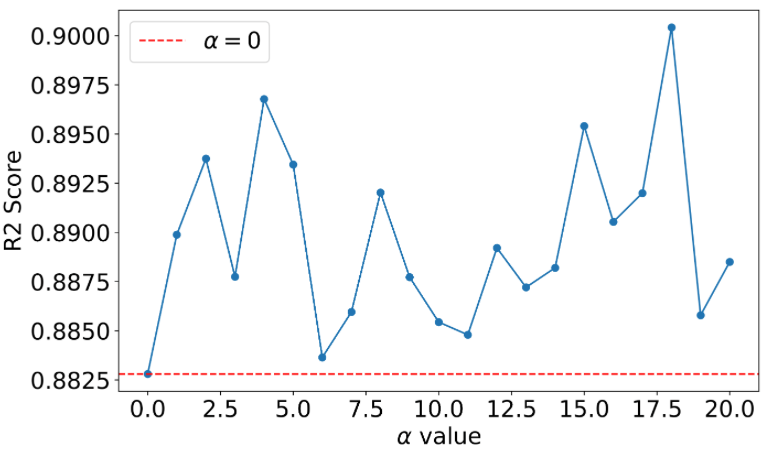}
    \caption{Plot of the \(\text{R}^2\) scores for varying \(\alpha\) values for the PV power dataset. The blue curve represents the \(\text{R}^2\) scores obtained for each \(\alpha\) value, with the red dashed line indicating the baseline performance at \(\alpha = 0\). The fluctuations illustrate how different \(\alpha\) values impact model performance, with all non-zero \(\alpha\) values surpassing the baseline score.}
    \label{fig:Results_pvpower}
\end{figure}

Based on the (\(\text{R}^2\) score results presented in Figure \ref{fig:Results_pvpower}, we observe that the model generally demonstrates improved performance when a non-zero \(\alpha\) value is introduced, as compared to the baseline scenario where \(\alpha = 0\) (indicated by the red dashed line). The \(\text{R}^2\) scores fluctuate as \(\alpha\) varies, all non-zero \(\alpha\) values yield scores higher than the baseline, with peaks observed at \(\alpha = 5\) and \(\alpha = 18\). These findings suggest that incorporating a non-zero \(\alpha\) value can enhance the model's ability to capture the underlying patterns in the data for regression tasks, further extending the benefits observed previously in classification tasks.

\subsection{Neural network example}
\label{sec: Example with neural network model}

Lastly, we wanted to explore if we could apply this concept to classical neural networks. For this, we utilized a dataset related to water potability \cite{noauthor_water_nodate}, which includes features such as pH, Hardness, Solids, and so on. The target variable, Potability, is a binary classification label indicating whether the water is potable or not.

After preparing the dataset with a standard classical preprocessing process, we constructed the neural network. 
The architecture of the network consisted of three fully connected dense layers. The input layer, corresponding to the nine input features, was connected to the first hidden layer comprising 36 neurons. The first hidden layer was followed by a dropout layer with a 50\% dropout rate to prevent overfitting. The second hidden layer also consisted of 36 neurons and was followed by another dropout layer with a 50\% dropout rate. The output layer consisted of a single neuron, which was used to produce the final binary classification output.
The ReLU activation function was applied to the outputs of both hidden layers to introduce non-linearity into the model. The final output used a sigmoid activation function to convert the output into a probability score suitable for binary classification.

We conducted experiments to evaluate the model's performance, comparing models with the parameter \(\alpha\) set to zero against models with non-zero \(\alpha\) values. Since classical learning rate are usually an order of magnitude smaller compared to quantum learning rate we had to increase the $\alpha$ value the order of magnitude by one. The key metrics were the speed of convergence, measured by the number of steps required to reach a stable loss, and the test accuracy. Early stopping was implemented to halt training if no improvement in validation accuracy was observed for 50 epochs.

For each value of \(\alpha\), including \(\alpha = 0\), we tested different learning rates to assess their impact on model performance. In total, we tested 20 different learning rates, ranging from 0.0001 to 0.4, to assess their impact on model performance.
The results were averaged across multiple runs to account for variability in training. The models with \(\alpha = 0\) served as the baseline, providing a benchmark for evaluating the impact of regularization.

Our experiments revealed that only the model with \(\alpha = 20\) showed an improvement in test accuracy, increasing slightly from approximately 66\% to 67\% compared to the baseline model with \(\alpha = 0\). Other models with non-zero \(\alpha\) values demonstrated lower test accuracy, indicating that increasing the $\alpha$ parameter did not consistently enhance generalization.

\begin{figure}[ht]
    \centering
    \includegraphics[width=\columnwidth]{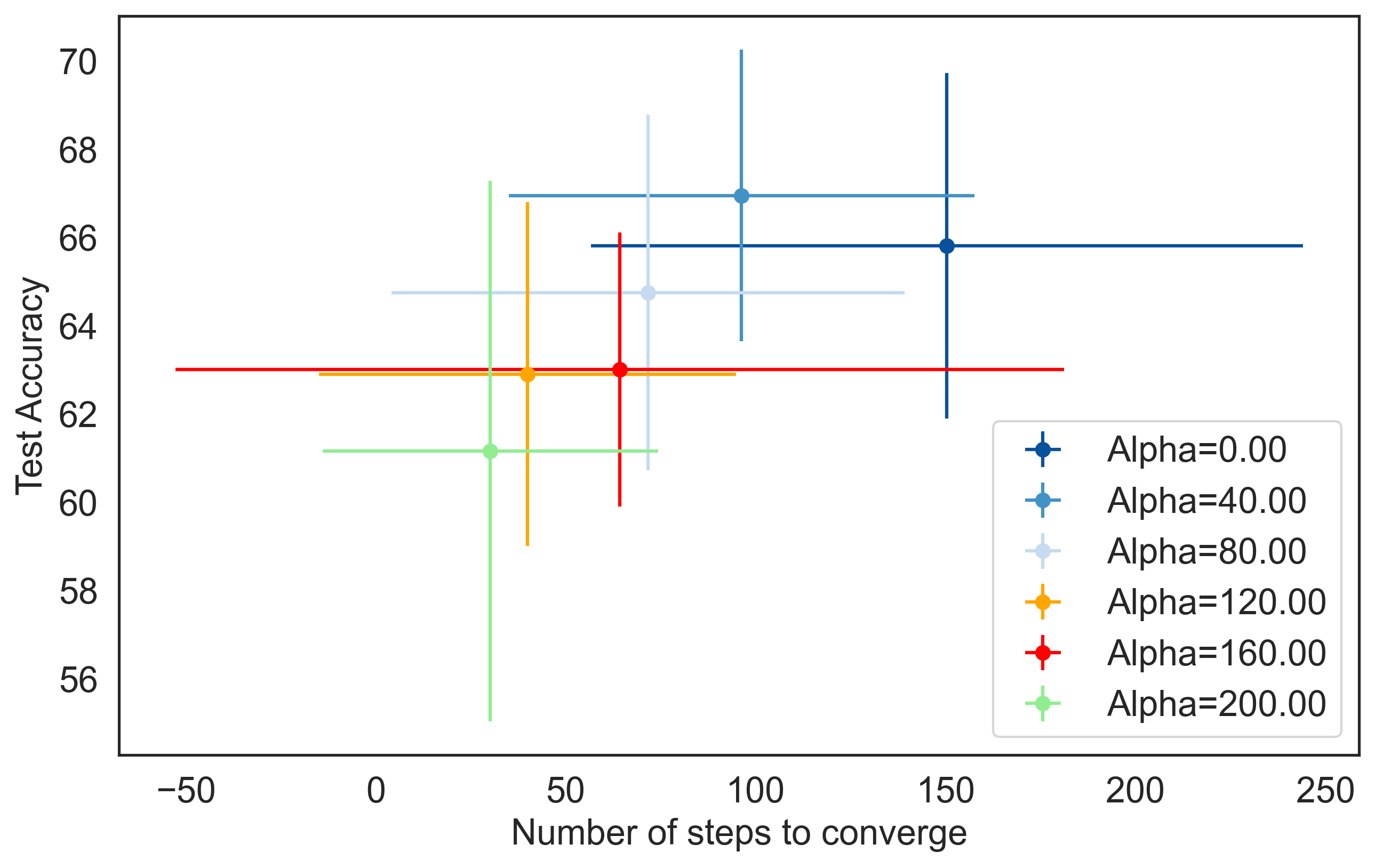}
    \caption{Plot of validation accuracy against the number of steps to converge for varying \(\alpha\) values. Each point represents the average performance for a given \(\alpha\) value, with error bars indicating the standard deviation across multiple runs. The colors correspond to different \(\alpha\) values, ranging from \(\alpha = 0.00\) (blue) to \(\alpha = 200.00\) (brown). This plot illustrates how different $\alpha$ values impact both the speed of convergence and the final accuracy.}
    \label{fig:Results_MI_classical}
\end{figure}

In addition, as shown in Figure~\ref{fig:Results_MI_classical}, models with non-zero \(\alpha\) values generally required fewer steps to converge, suggesting faster training times. 
However, this did not necessarily translate into improved accuracy. 
These findings, though preliminary, suggest that while regularization may accelerate convergence, it does not guarantee better model performance. 
Further analysis with a broader range of \(\alpha\) values and additional metrics would be needed to draw definitive conclusions.

\end{document}